\documentclass[twocolumn,a4paper,12pt]{article} 
\usepackage{epsfig}
\usepackage{graphicx}

\newcommand{\beq}{\begin{eqnarray}}
\newcommand{\eeq}{\end{eqnarray}}

\begin{document}
\pagestyle{plain}

%
%
\title{
{\Large \bf 
Identifying high power breakdowns in accelerating structures with acoustic sensors
}
}
 
\author{
{\bf 
Nicolas Delerue
} \\ 
GLC Group, KEK\\
Talk given at the 9th Accelerator and Particle Physics Institute
}

\date{
APPI (Japan), February 2004
}

\twocolumn[
\vspace*{-2cm}
\maketitle

\vspace*{-1cm}
%
\begin{center}
{\bf 
To increase the energy of Particle Accelarators to reach the requirements of the Linear Collider or of the Super KEK-B factory, new accelerating structures with a higher accelerating gradient need to be developped. These accelerating structures are often affected by high power breakdowns. Knowing the exact location where these breakdowns occured helps to redesign the structure and improve its performances.\\ This can be done with acoustic sensors.
} \\ 
\end{center}
]

\vspace*{.5cm}

%
\section{Accelerating strucures R\&D}
%

\begin{figure}[hbtp]
\centering\epsfig{file=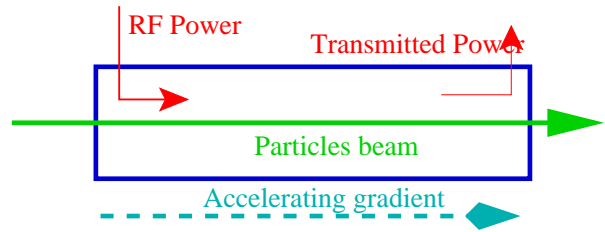,width=8.cm}
\caption{Scheme of an accelerating structure. The electromagnetic wave
(red arrows) is inserted at the left of the structure and travels to
  the right, thus creating an accelerating gradient that will
  accelerate the particles beam. At the end of the structure the
  transmitted power
  is extracted and can be measured.}
\label{fig:accStruct} 
\end{figure}

Accelerating structure are used in (mainly in LINACs) to accelerates particles. Usually a very high frequency (Radiofrequency) electromagnetic pulse is feed into the structure.  This electromagnetic pulse will create, in the structure, an accelerating gradient that will accelerate the particles beam (see figure~\ref{fig:accStruct}).

%
%

\begin{figure}[hbt]
\centering\includegraphics[width=0.50\textwidth,angle=0]{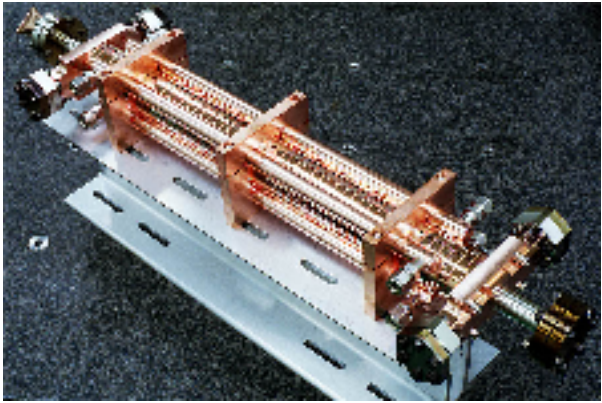}
\caption{A x-band accelerating structure.}
\label{fig:accStructXband} 
\end{figure}

At KEK-B the typical power feed into a structure is 40~MW at 3~GHz (S-band) whereas the structure that should be used in the Linear Collider~\cite{unknown:2003mg} (if warm technology is chosen) will be feed with 75~MW at a frequency of 11~GHz (X-band). A picture of an X-band accelerating structure is shown on figure~\ref{fig:accStructXband}.

As the accelerating gradient depends to the power that was feed into the structure, to increase the accelerating gradient (and reach higher energies) one needs to increase the power of the pulse.

%
%

For example, for the upgrade of KEK-B, a higher accelerating gradient is needed as this will allow to reach a higher luminosity. Thus the current S-band accelerating structures will be replaced by C-band (6~MHz) accelerating structures operating with a higher power input (More details on the KEK-B upgrade and the C-band structures R\&D can be found in Sugimura Takeshi's contribution to this APPI workshop).

For the Linear Collider the required energy makes it necessary to use X-band structures. Such structures have never been used in an accelerator before before and thus their design is new.

%
\subsection{High power breakdowns}
%

\begin{figure}[hbt]
\centering\epsfig{file=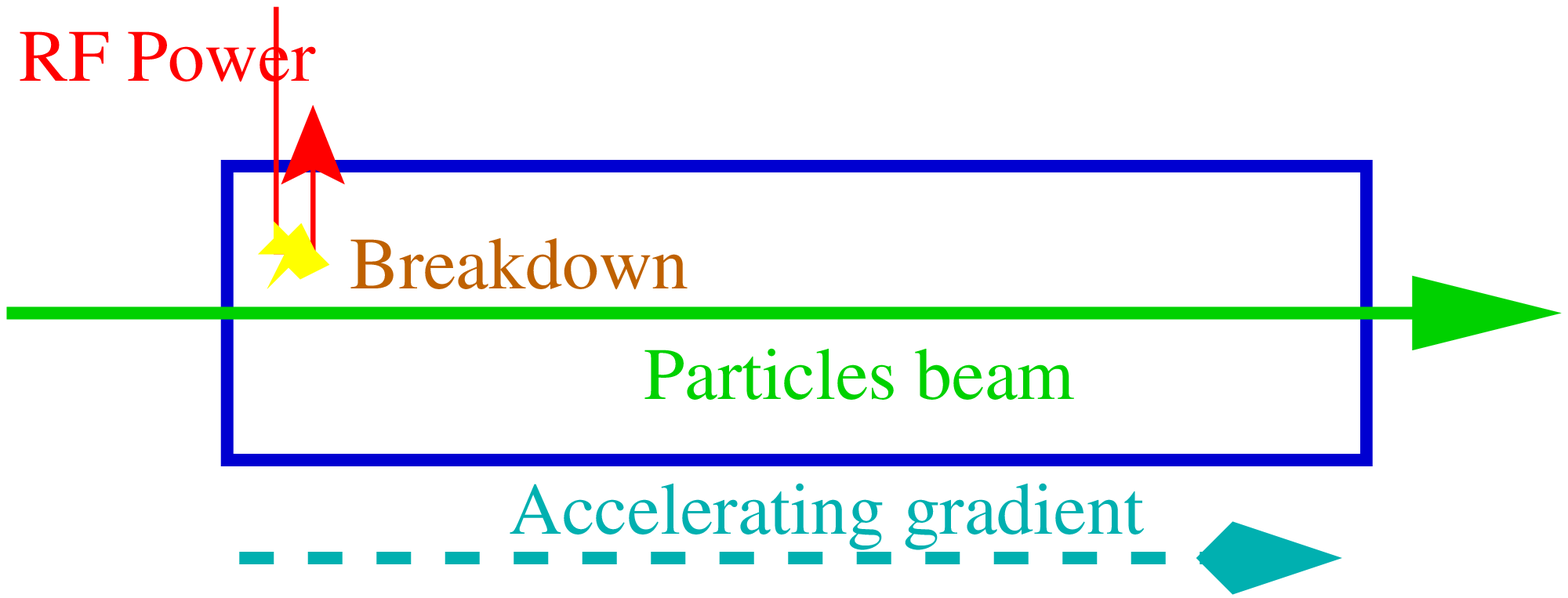,width=8.cm}
\centering\epsfig{file=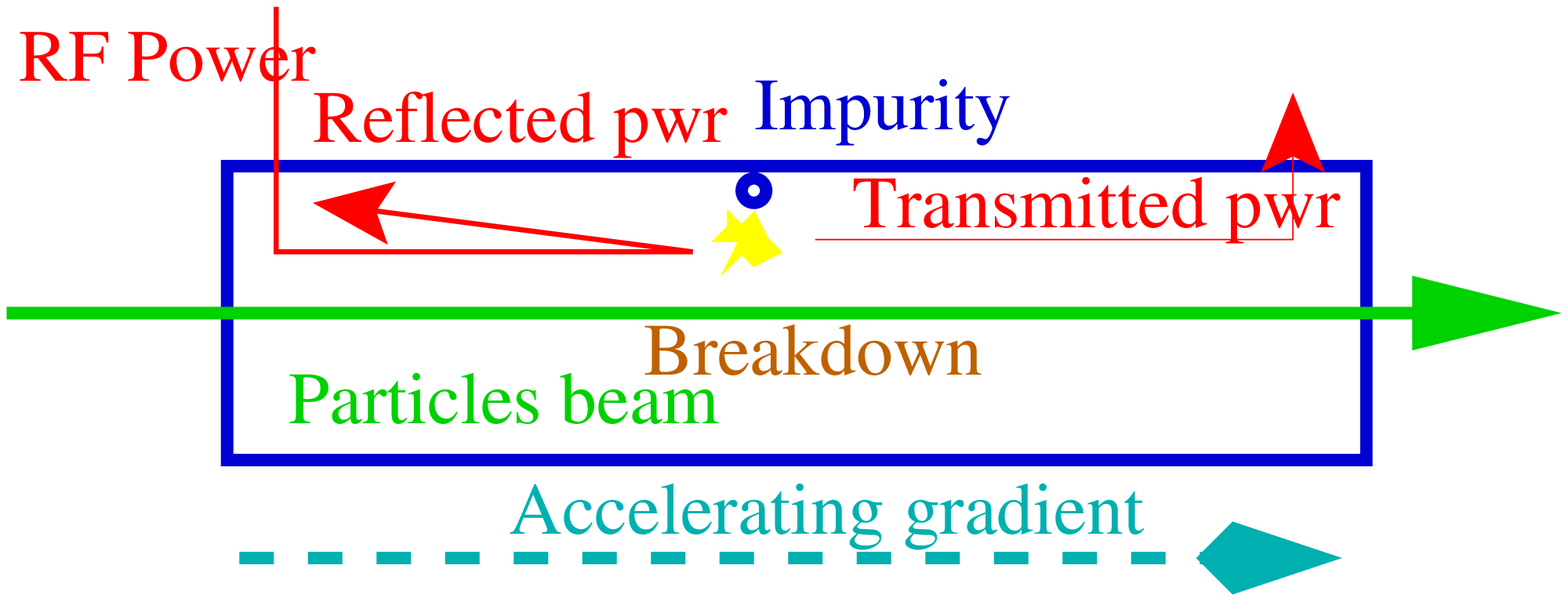,width=8.cm}\\
\caption{Breakdowns in an accelerating structure.
Breakdowns can occur either at the input or output of the structure
(top figure) or somewhere else in the structure, for example near an impurity at the
 surface of the structure (bottom figure).
The timing with which power will be transmitted and reflected depends
on the location of the breakdown.}
\label{fig:accStructBkdn} 
\end{figure}

When the energy concentrated on a given point of the structure becomes
too high a spark will occur. This can happen at a location where, by
design, the energy density is very high, like for example the input
coupler of the structure (see figure~\ref{fig:accStructBkdn} top) but
this can also happen at a location where there is an impurity at the
surface of the metal forming the structure (see
figure~\ref{fig:accStructBkdn} bottom).

%
%

A proper identification of the location where the breakdowns occur helps to redesign (in future structures) the location where the power concentrate.

%
\subsection{Breakdowns identification}
%

There are several different methods to locate where a breakdown occured. 

\begin{itemize}
\item As shown on figure~\ref{fig:accStructBkdn} when the breakdown occurs the incoming electromagnetic wave is reflected back to the input of the structure. If the breakdown occurs in the middle of the electromagnetic wave, then part of the wave will be reflected whereas the part that had already passed the location of the breakdown will be transmitted. Thus by measuring the shape of the reflected and trasmitted waves, it is possible to deduce the location and the time (with respect to the begining of the wave) at which the breakdown occured. As the electromagnetic wave is often deformed, this method has a limited precision.
\item When the breakdown occurs, noise is produced. By placing accoustic sensors (microphones) near the structure it is possible detect this noise. If the accoustic sensors are located regularly along the structure it is possible to identify which sensor received the signal first and thus near which sensor the breakdown occured.
\item The breakdowns also produces X-rays that can be detected with conventionnal X-Ray detectors.
\item Other methods are exist but are often more expensive or less easy to use than the one mentionned above.
\end{itemize}

%
\section{Breakdowns studies for the C-Band structures}
%
The C-band accelerating structure R\&D done at KEK aims to have a suitable structure design for the KEK-B upgrade. This work is conducted by the KEK Linac Upgrade group conducted by Kamitani Takuya (More details on this work can be found in Sugimura Takashi's presentation\cite{cband_talk} during this workshop).

\begin{figure}[hbtp]
\centering\epsfig{file=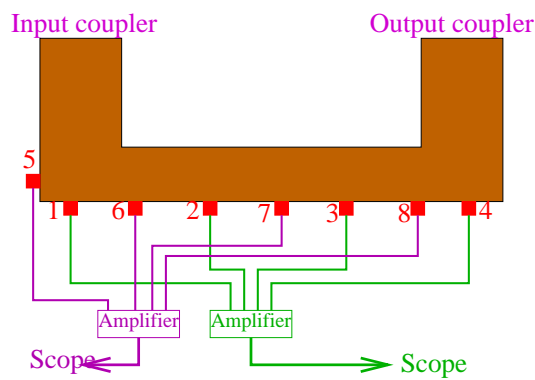,width=7.cm}
\caption{Layout of the 8 acoustic sensors installed on the C-band structure tested at KEK's main Linac.}
\label{fig:cband_layout} 
\end{figure}

\begin{figure}[hbtp]
\centering\epsfig{file=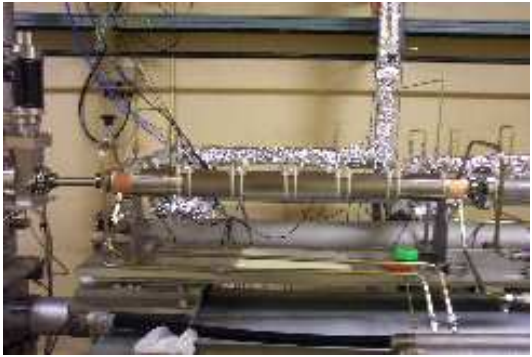,width=7.cm}
\caption{The C-band structure tested at KEK's main Linac. The acoustic sensors can be seen on top of the structure.}
\label{fig:cband_wide_view} 
\end{figure}

\begin{figure}[hbtp]
\centering\epsfig{file=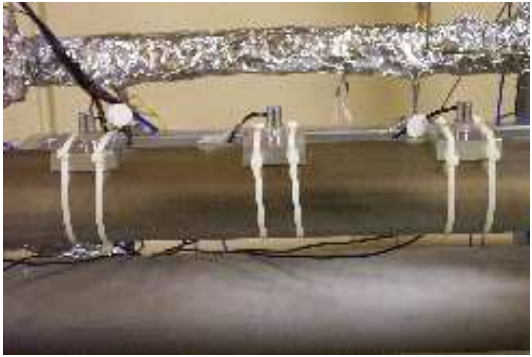,width=7.cm}
\caption{Zoom on 3 sensors installed on top of the C-band structure tested at KEK's main Linac.}
\label{fig:cband_zoom_view} 
\end{figure}

There is one accelerating structure currently tested at the KEK main Linac. It has been equipped with 4 acoustic sensors, to which 4 extra sensors where added later as shown on figures~\ref{fig:cband_layout}, ~\ref{fig:cband_wide_view} and ~\ref{fig:cband_zoom_view}. Additionaly after each breakdown informations about the RF wave are stored and available for analysis.

%
%
%

%
%
%

%
%
%

%
%
\begin{figure}[hbtp]
\centering\epsfig{file=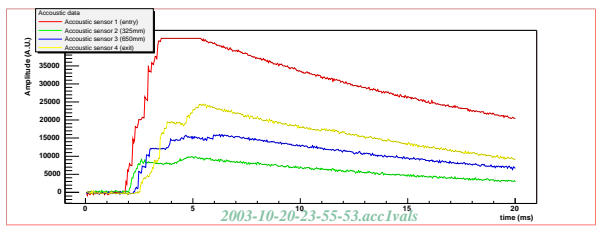,width=8cm}
\caption{Example of acoustic data recorded during a breakdown at the
  C-band test stand. Each  line corresponds to the noise recorded by a
  differenr sensor.}
\label{fig:bkdn_acc1} 
\end{figure}

\subsection{Acoustic data}

On figure~\ref{fig:bkdn_acc1}  one can see an example of breakdown as recorded by the acoustic sensors. The signal is huge and thus easily distinguishable from the background noise. But extracting the breakdown location from this signal is not so easy. One could decide that the sensor the closest from the breakdown will receive the signal earlier (the travel path is shorter) or the more intense (the attenuation should be smaller).

\begin{figure}[hbt]
\centering\epsfig{file=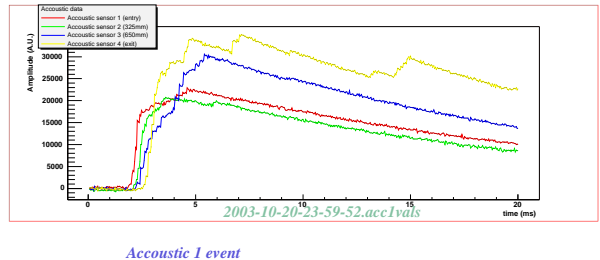,width=8.cm}
\caption{Example of acoustic data recorded during a breakdown. On this event one can see that the sensor receiving the signal first is not the one that receives the most intense signal.}
\label{fig:bkdn_acc2} 
\end{figure}

Unfortunately on some events such as the one shown on figure~\ref{fig:bkdn_acc2} the sensor receiving the signal first is not the one that receives the most intense signal. Furthermore the order in which the sensors receive the signal will not always have a physical interpretation. Thus a double algorithm has been used to locate the breakdown: one is based on the time at which each sensor has received the signal and the other one is based on the intensity of the signal. Attempts to use either intensity or timing information to locate the breakdown between two sensors have not been successful.

%
%

\begin{figure}[hbtp]
\twocolumn[
\centering\epsfig{file=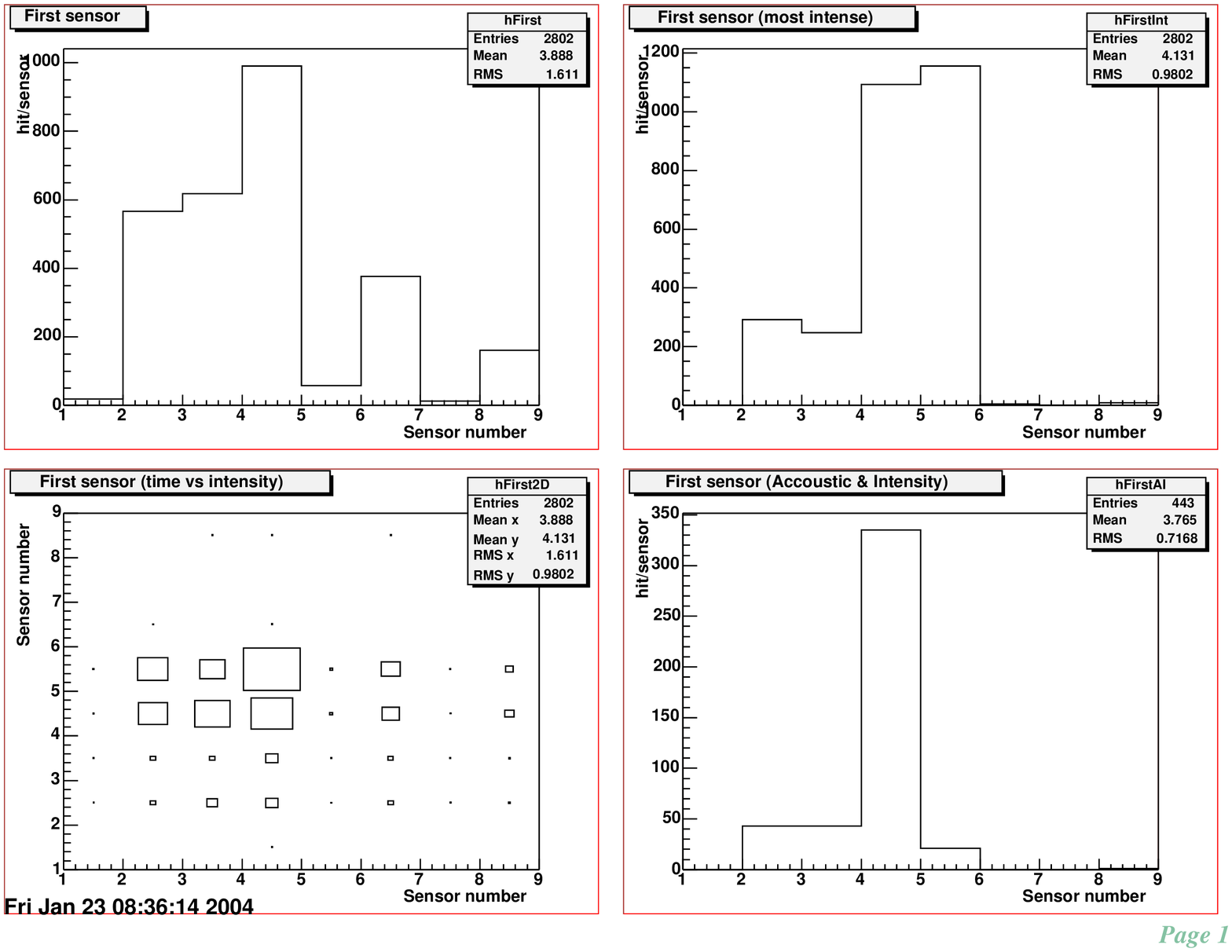,width=12cm,angle=270}
\caption{Analysis of the breakdowns recorded with the acoustic
  sensors. The upper left plot shows for each sensor the number of
  breakdowns where this sensor received the noise first. The upper
  right plot shows for each sensor the number of
  breakdowns where this sensor received the most intense signal. The
  lower left plot shows the correlation between the two upper plots,
  the horizontal axis corresponding to the sensor that received the
  noise first and the vertical axis to the sensor that received the
  most intense signal. The lower right plot shows for each sensor the
  number of breakdowns where this sensor received the signal first and
  also received the most intense signal.}
\label{fig:accStats} 
\vspace{0.5cm}
]
\end{figure}

The number of breakdowns recorded by each sensor can be seen on figure~\ref{fig:accStats}. The upper left plot shows the number of breakdowns that have reached a sensor first, the upper right shos the number of breakdons that were the most intense for a given sensor. The lower left plot shows the correlations between the to upper plots and the lower right plots shows when the number of breakdowns when timing and intensity information agree (golden events).

%
%

From this figure one can see that the breakdowns occur in the forward region of the structure as sensor 2 to 4 record the most intense signal. By looking at the golden events one can see that these events seems to be located inside the structure rather than at its edge as it is sensor 4 that records the more golden events.

\subsection{Electromagnetic wave data}

%
%
\begin{figure}[hbt]
\centering\epsfig{file=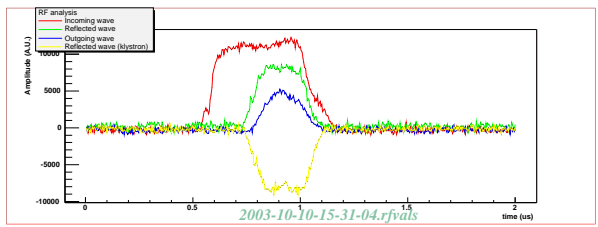,width=8.cm}
\caption{Example of RF data recorded during a breakdown. The red line
  corresponds to the incoming RF wave, the green line to the reflected
  RF wave, the blue line to the outgoing wave and the yellow line to
  the reflected wave at the klystron..}
\label{fig:bkdn_rf} 
\end{figure}

The analysis of the electromagnetic waves is more straightforward. The figure~\ref{fig:bkdn_rf} shows an example of electromagnetic wave measurement during an event where a breakdown occured. 
The difference between the time at which the outgoing wave (in blue) finishes and the time at which the reflected wave (in green) reaches the input of the structure provides information on the location of the breakdown: if the breakdown occured close from the output of the structure the outgoing wave will have to travel a much shorter distance than the reflected wave, thus the time difference between the two will be negative. On the other hand, if the breakdown occurs near the input coupler the reflected wave will have a short travel distance whereas the outgoing wave will have to travel a much longer path, thus the difference will be positive.

The comparison between the length of the incoming electromagnetic wave and the length of the outgoing wave gives the time at which the breakdown occured.

To measure the resolution of this analysis, it is possible to look at the difference between the time at which the incoming wave is recorded and the time at which the outgoing wave is recorded. This difference corresponds to the time an electromagnetic wave needs to travel the length of the structure and should be constant. The distribtion of this variable gives the resolution of the other variables. 

\begin{figure}[htbp]
\centering\epsfig{file=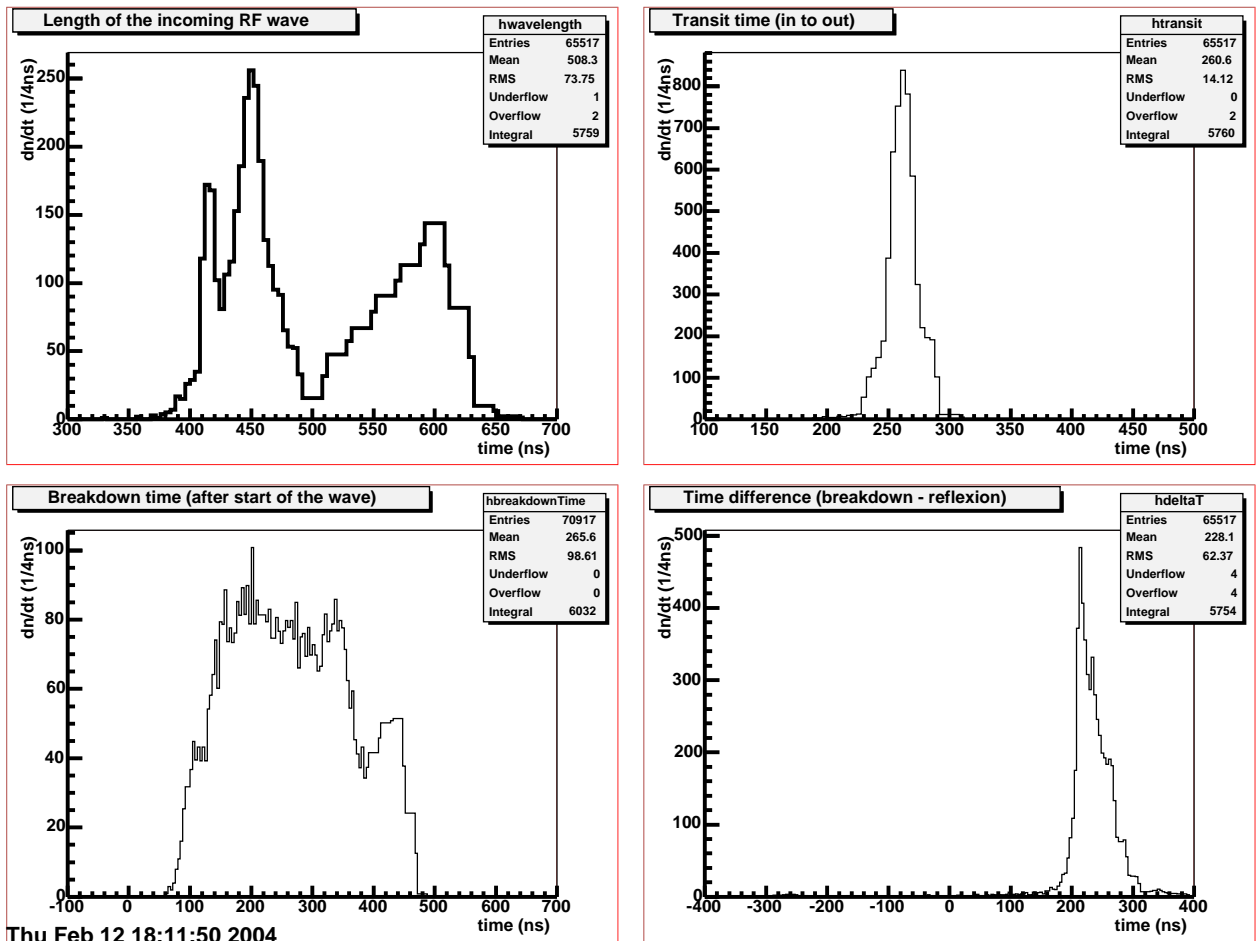,width=12.cm,angle=270}
\vspace*{0.5cm}
\caption{Analysis of the electromagnetic waves information
  recorded with each breakdown. The upper left figure shows the total
  length of the incoming RF wave. The upper right plot shows the
  transit time from the input to the output of the coupler. As this
  time is a constant the width of the peak gives the resolution of
  our measurement. The lower left plot shows the breakdown time after
  the start of the wave. The lower right plot gives the difference
  between the time at which the outgoing and reflected waves were
  recorded. Positive values correspond to breakdowns near the input kicker.}
\label{fig:rfStats} 
\end{figure}

The analysis of the electromagnetic waves information is shown on
figure~\ref{fig:rfStats}. From this analysis one can see that there is no specific time at which the breakdown occurs but that almost all breakdowns occur near the input coupler.

\subsection{Correlations between electromagnetic and RF wave data}

\begin{figure}[htbp]
\centering\epsfig{file=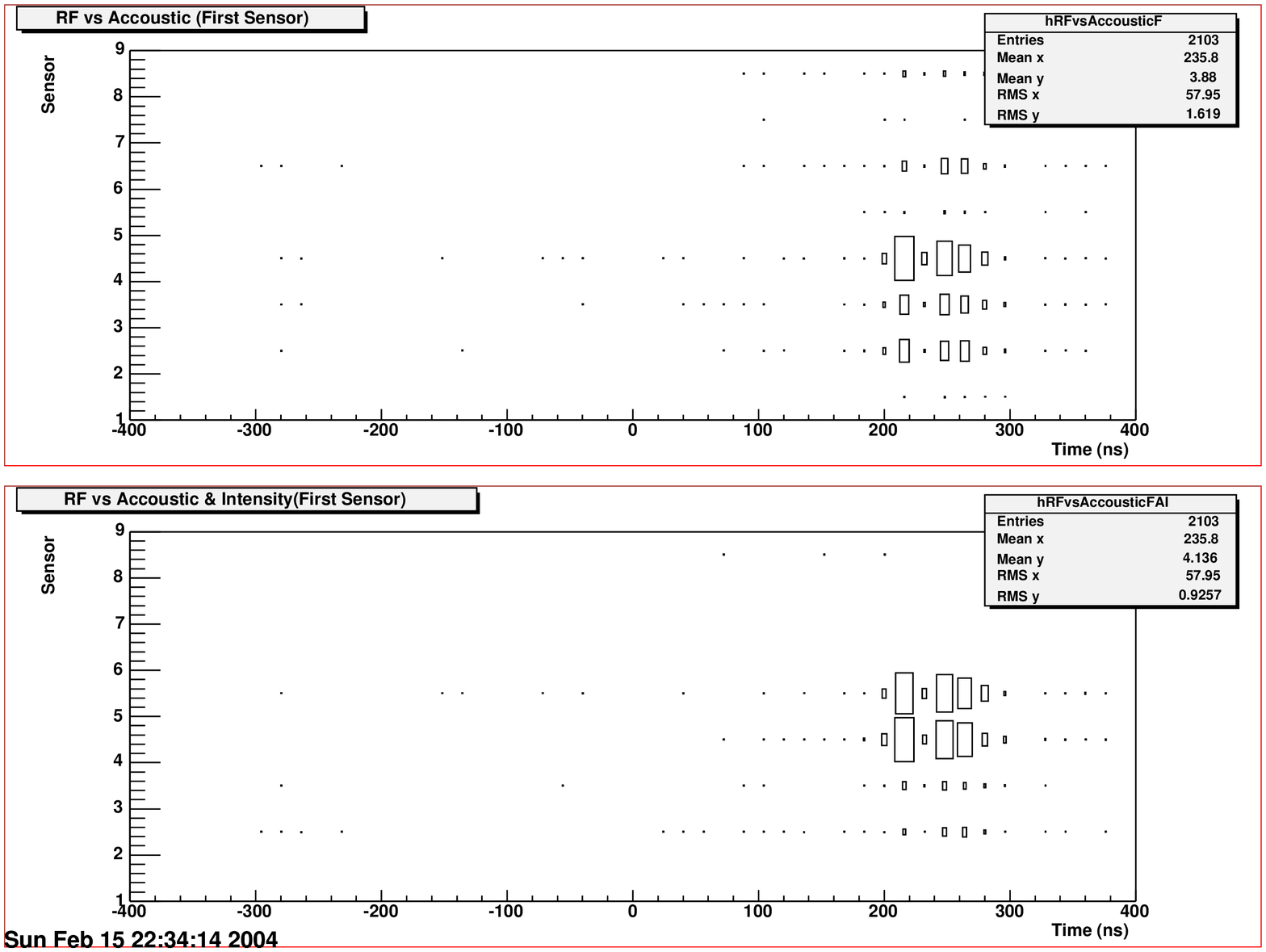,width=12.cm,angle=270}
\vspace*{0.5cm}
\caption{Correlations between the acoustic sensors analysis (vertical scale) and the
  electromagnetic waves analysis (horizontal scale).}
\label{fig:CorelStats} 
\end{figure}

The correlation between the information collected with the acoustic sensors and the electromagnetic waves analysis has been studied. This correlation can be seen on figure~\ref{fig:CorelStats}.

\onecolumn
%
%
%
%
%
\twocolumn

%
\section{Breakdowns studies for the X-Band structures}
%

The X-band accelerating structure R\&D done at KEK is part of the worldwide effort to build a X-band Linear Collider (GLC, Global Linear Collier) able to reach a center of mass energy of 1 TeV. This work is conducted at the GLCTA (GLC Test Accelerator) headed by Higo Toshiyasu.

A new test stand has recently been installed at the GLCTA and was under commisionning until March 2004. Thus acoustic sensors were used to detect breakdowns occuring in the waveguide and not in the structure itself. The acoustic sensors used at the GLCTA are read by VME modules. The sensors and the modules have been provided by SLAC where a similar system is already in use\cite{Nelson:2003ad}.

In a first step only 16 sensors have been deployed but in a later step up to 400 sensors will be installed on the structure once the test stand will be fully commissionned.

The VME data acquisition software and the data analysis software have both been written at KEK.

\begin{figure}[hbtp]
\centering\epsfig{file=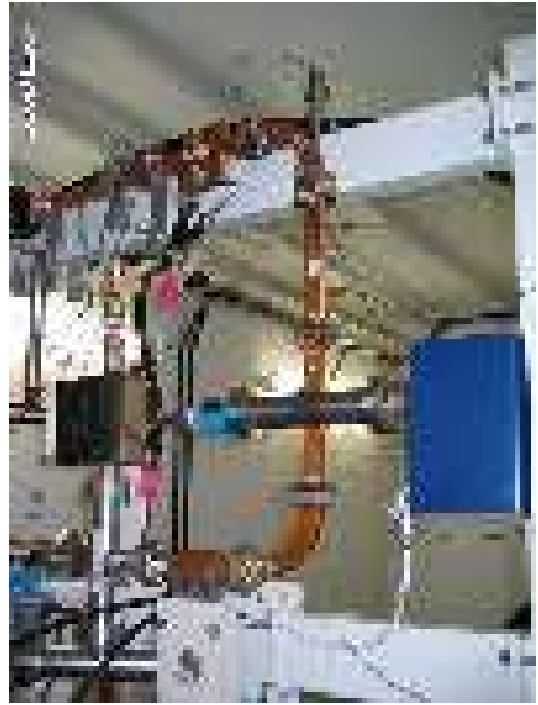,width=7.cm}
\caption{A waveguide of the X-band test stand equipped with acoustic sensors.}
\label{fig:xband_wide_view} 
\end{figure}

\begin{figure}[hbtp]
\centering\epsfig{file=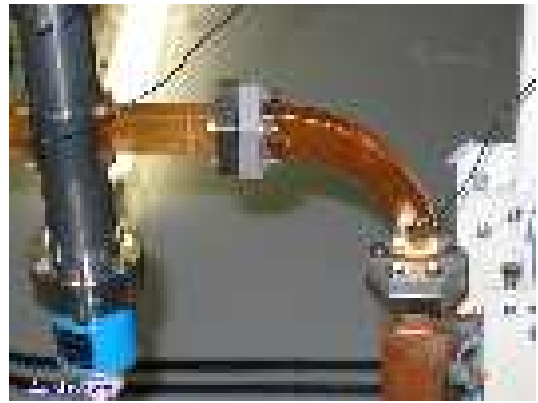,width=7.cm}
\caption{A piece of waveguide of the X-band test stand with an
  acoustic sensor.}
\label{fig:xband_close_view} 
\end{figure}

A usual glue was used to stick the acoustic sensors on the waveguides
at the location know to be the most likely to produce
breakdowns. Acoustic sensors glued on the waveguides can be seen on figures~\ref{fig:xband_wide_view} and~\ref{fig:xband_close_view}.

%
%
%

%
%

\subsection{Acoustic data analysis}

\begin{figure}[p]
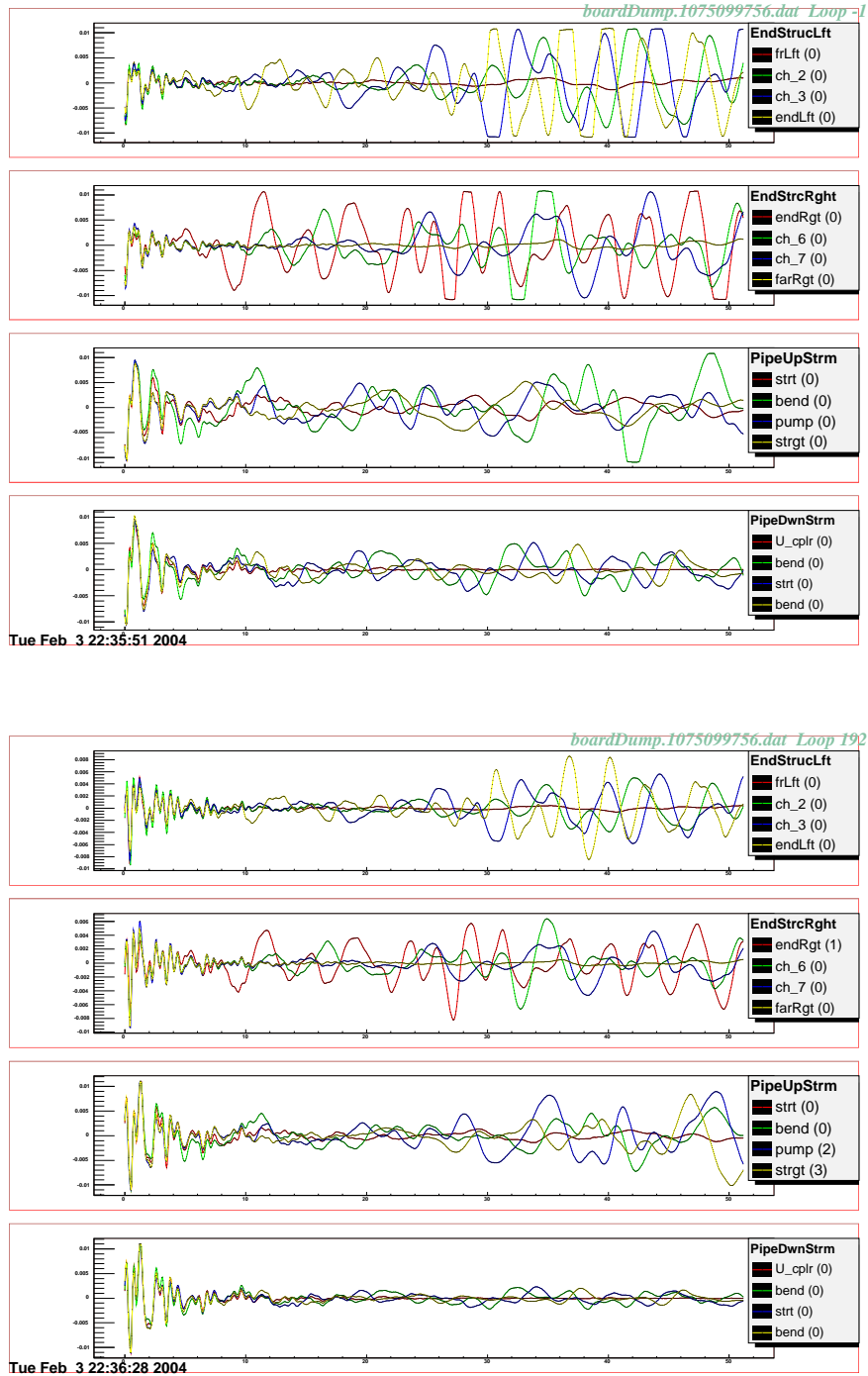

\centering\epsfig{file=boardDump.1075099756.dat_boardDump.ps_page_1,angle=270,width=5.5cm}\\
\vspace{1.5cm}
\centering\epsfig{file=boardDump.1075099756.dat_boardDump.ps_page_6,angle=270,width=5.5cm}
\caption{X-band events. On the top figure there is a breakdown on the
  4 upper channels (upper plot) whereas on the bottom figure the noise
  recorded does not correspond to an actual breakdown.}
\label{fig:xband_event} 
\end{figure}

The figure~\ref{fig:xband_event} page~\pageref{fig:xband_event} (top)  shows a typical event as recorded by
the acoustic sensors at the X-band test stand.

On this event one can see that even before the breakdown there is a
lot of noise recorded by the sensor, especially the klystron noise at
the beginning of the record. The figure~\ref{fig:xband_event} (bottom)
shows an example of non breakdown event with high intensity noise.

Thus to identify a breakdown it is necessary to keep balance between
the use of a low threshold which would catch all breakdowns but would also
catch non breakdown noise and a higher threshold that would not catch
noise but might miss some of the breakdowns.

To address this issue 3 different level of breakdowns have been
defined:
\begin{itemize} 
\item A ``simple'' level, where all events where the signal passes a given
  threshold will be considered as a breakdown.
\item A ``long'' level, where all events where the signal passes a given
  threshold and remains above this threshold for a given time will be considered as a breakdown.
\item An ``extended'' level, where all events where the signal passes a given
  threshold and remains above this threshold for a given (extended) time will be considered as a breakdown.
\end{itemize} 

Furthermore most of the systematic component of the noise (such as
klystron noise) can be removed by substracting to the recorded noise
the average noise recorded during 200 recent events.

\begin{figure}[hbt]
\centering
\epsfig{file=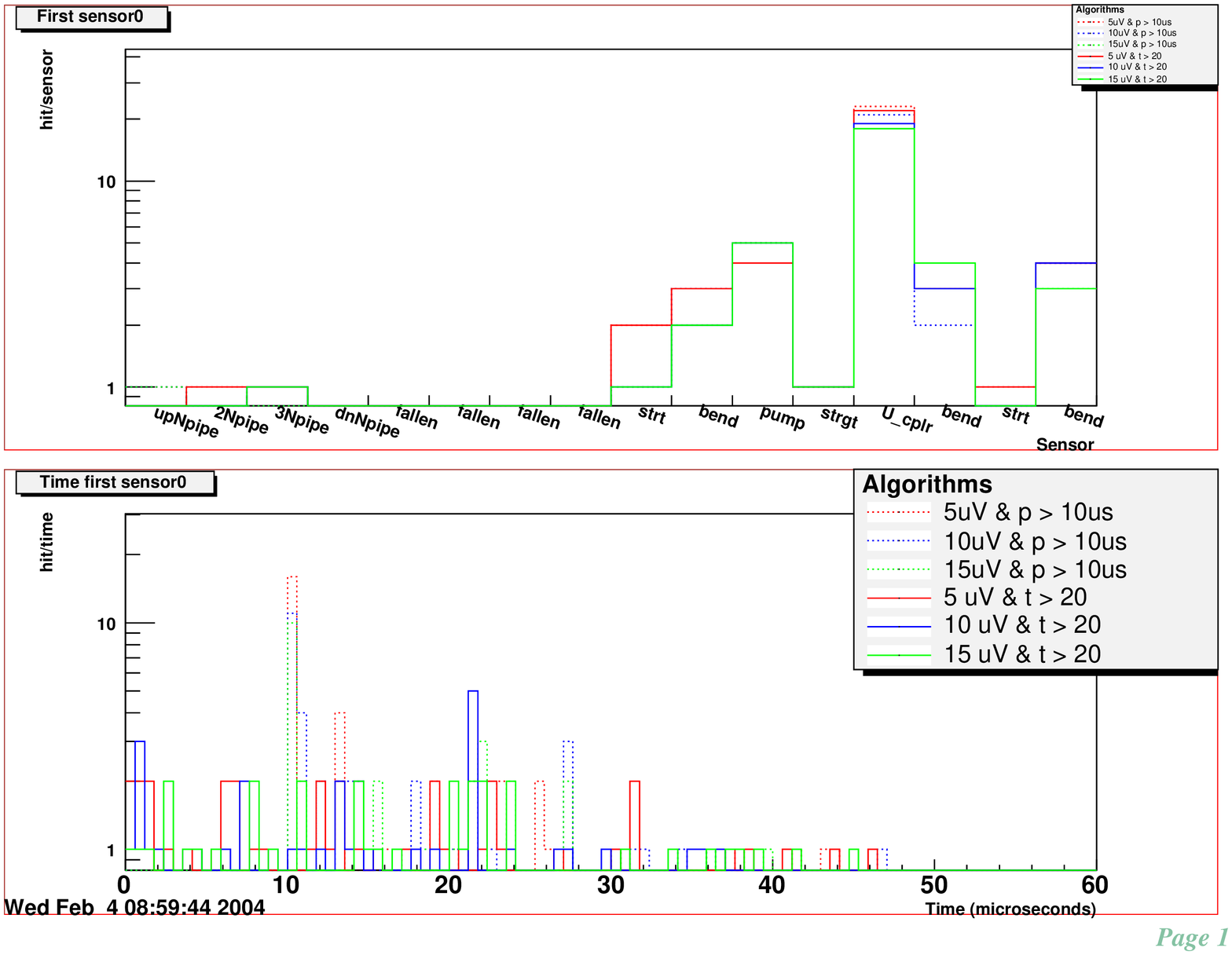,angle=270,width=6cm}\\
\epsfig{file=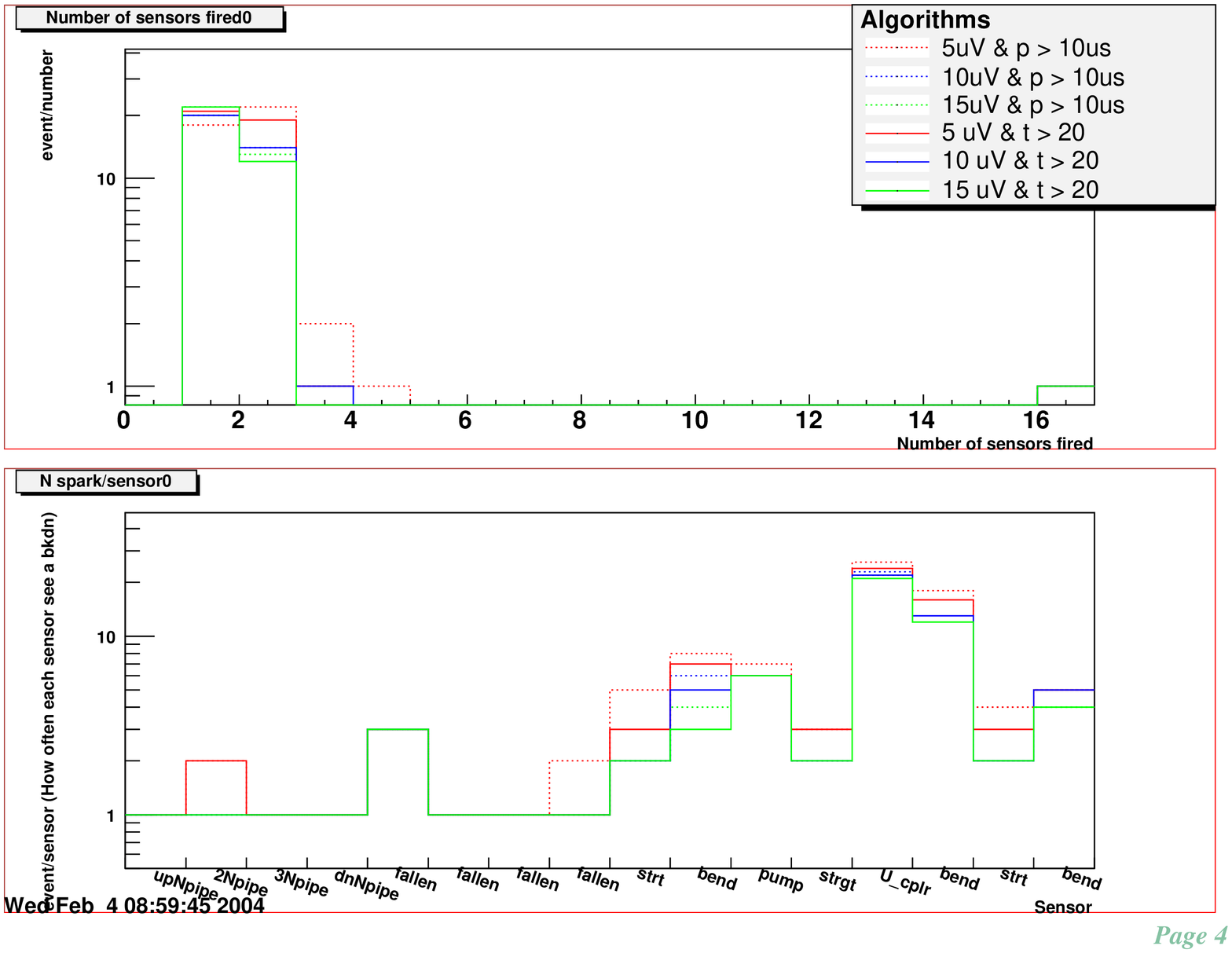,angle=270,width=6cm}\\
\caption{X-band analysis. All sensors used for this analysis were installed on waveguide
  components. The different colors correspond to different breakdown
  threshold. Dashed line correspond to simple breakdown and plain line
  to long breakdowns.}
\label{fig:xband_analysis_01} 
\end{figure}

\begin{figure}[hbtp]
\centering
\epsfig{file=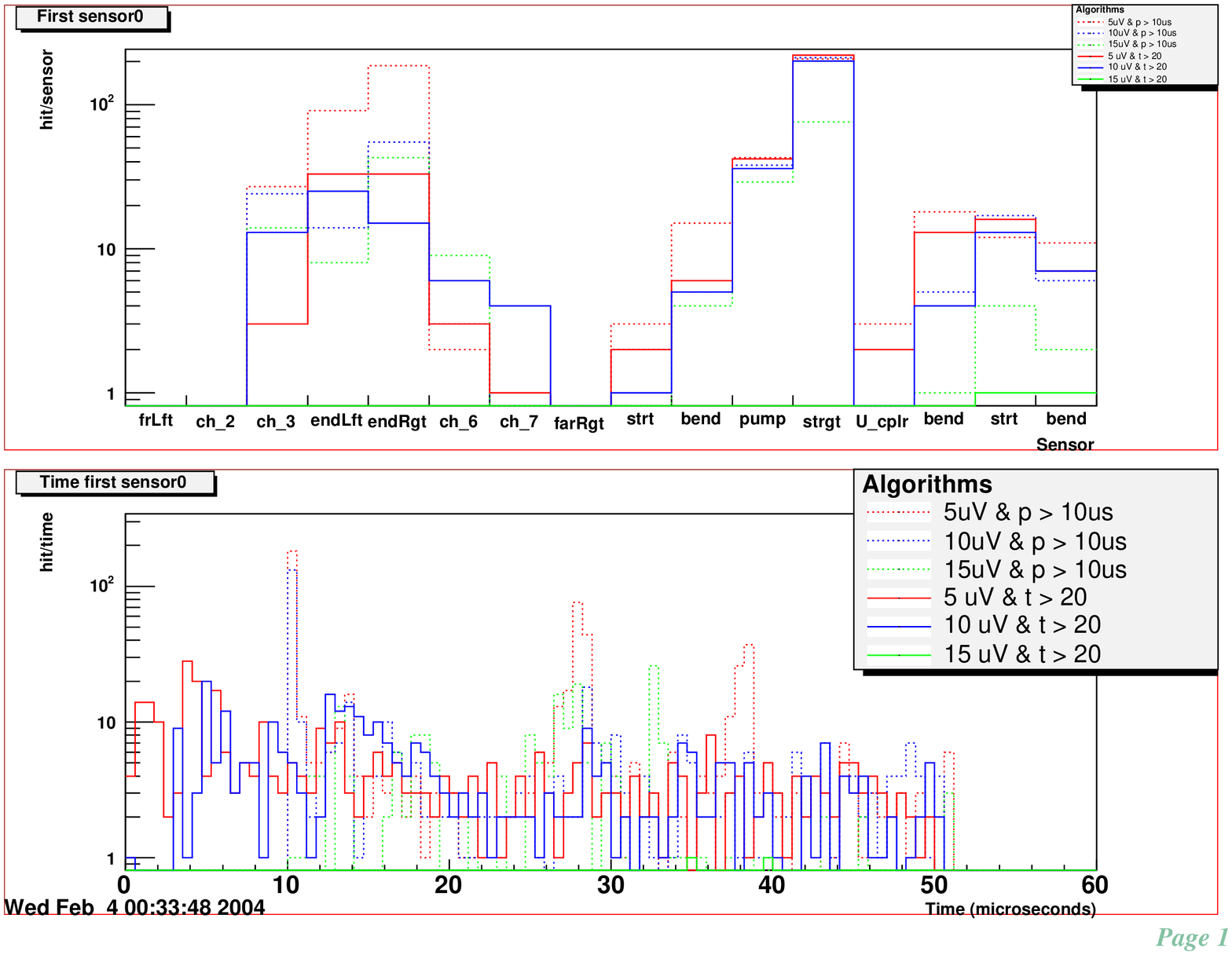,angle=270,width=5.8cm}\\
\epsfig{file=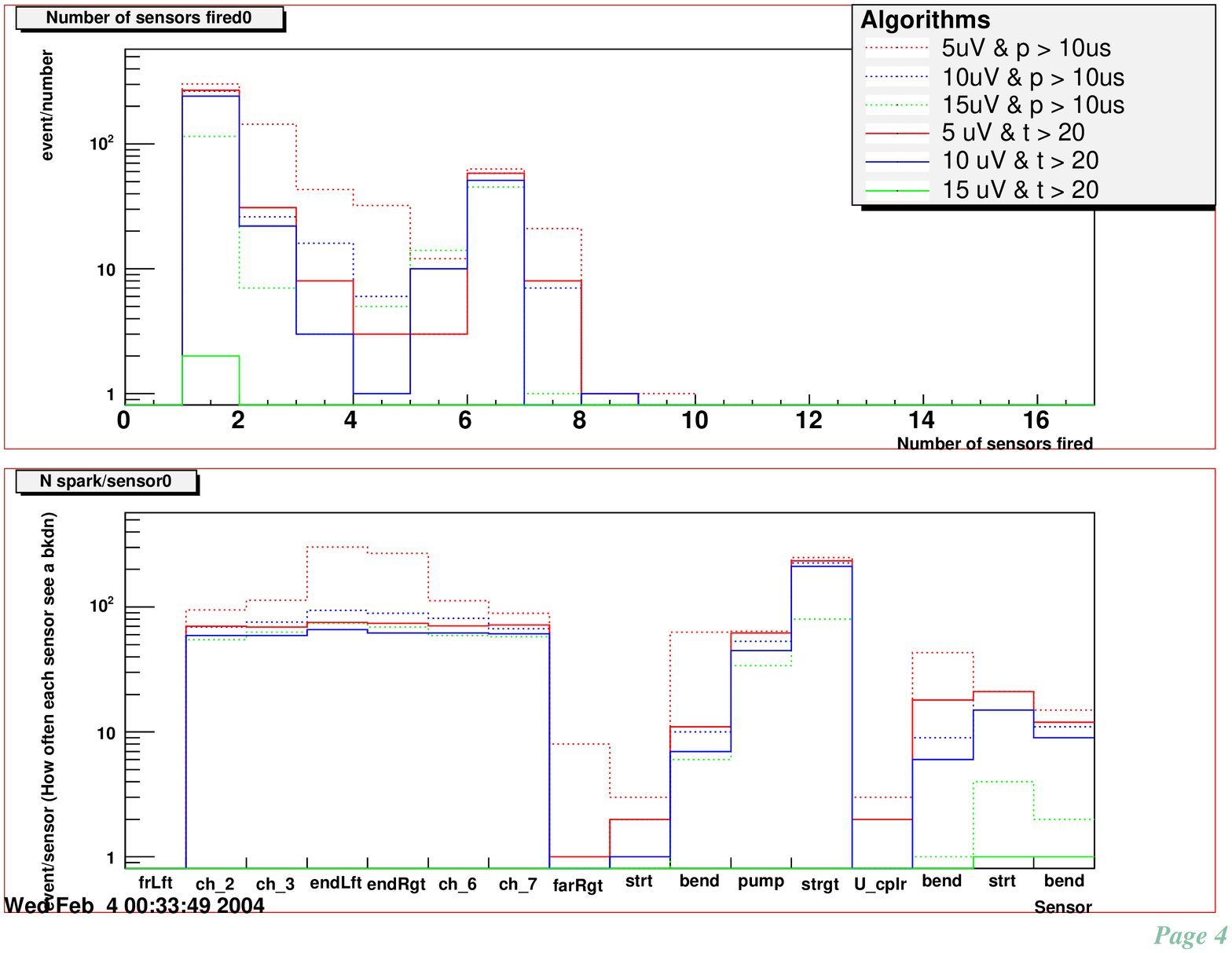,angle=270,width=5.8cm}
\caption{X-band analysis. Half of this sensors used for this analysis were installed on waveguide
  components and the other half were installed at the exit of the
  accelerating structure. The different colors correspond to different breakdown
  threshold. Dashed line correspond to simple breakdown and plain line
  to long breakdowns.}
\label{fig:xband_analysis_02} 
\end{figure}

The statistical repartition of the breakdowns following this
classification can be seen on figures~\ref{fig:xband_analysis_01}
and~\ref{fig:xband_analysis_02}.

These two figures correspond to different configuration of the
acoustic sensors. Figure~\ref{fig:xband_analysis_01} uses data where
all sensors were located on the waveguides. Some of them were located
on straight section of the pipes whereas some other were located on
bends or on a coupler. As one can expect most of the breakdowns where
recorded near the bent waveguides and near the coupler. Half of these
sensors were moved for the data used by
figure~\ref{fig:xband_analysis_02}. Those sensors have been relocated
at the end of the accelerating structure, another location where many
breakdowns (but also a lot of noise) were expected. The location of
the beakdowns on that figure is conform to the expectations.

\onecolumn
%
%
%

%
%
%

%
%

\clearpage


%
\section{Conclusion}
%

It has been verified that the information extracted from the acoustic
sensors is correlated with the information available from other
sources.

Acoustic sensors can provide useful information on the location of
breakdowns in an accelerator structure. They are very easy to install
and to move to survey different areas of the test stand. The
resolution that can be acheived is directly related to the number of sensors
available.

\bibliographystyle{myunsrt}
\bibliography{proceedings}

\begin{thebibliography}{1}

\bibitem{unknown:2003mg}
GLC project: Linear collider for TeV physics.
\newblock KEK-REPORT-2003-7.

\bibitem{cband_talk}
Sugimura Takashi.
\newblock Presentation at the APPI workshop, 2004.

\bibitem{Nelson:2003ad}
J.~Nelson et~al.
\newblock Use of acoustic emission to diagnose breakdown in accelerator RF
  structures.
\newblock Contributed to Particle Accelerator Conference (PAC 03), Portland,
  Oregon, 12-16 May 2003.

\end{thebibliography}

\end{document}